\documentclass[twocolumn,twocolappendix]{aastex701}

\usepackage{amsmath}
\usepackage{amssymb}
\usepackage{graphicx}
\usepackage{xcolor}
\let\tablenum\relax
\usepackage{siunitx}
\usepackage{xspace}
\usepackage{lettrine}

\input Zallman.fd

\newcommand{\GitHubURL}{https://github.com/lodubay/twoinfall}
\newcommand{\vice}{{\tt VICE}\xspace}

\newcommand{\osuaffil}{Department of Astronomy, The Ohio State University, 140 W. 18th Ave, Columbus OH 43210, USA}
\newcommand{\ccappaffil}{Center for Cosmology and AstroParticle Physics, The Ohio State University, 191 W. Woodruff Ave., Columbus OH 43210, USA}
\newcommand{\aFe}{[$\alpha$/Fe]\xspace}
\newcommand{\mathXH}{{\rm [X/H]}}
\newcommand{\mathOH}{{\rm [O/H]}}
\newcommand{\mathFeH}{{\rm [Fe/H]}}
\newcommand{\mathOFe}{{\rm [O/Fe]}}

\newcommand{\yZ}[1]{$y/Z_\odot=#1$}
\newcommand{\kpc}{\,{\rm kpc}}
\newcommand{\Myr}{\,{\rm Myr}}
\newcommand{\Gyr}{\,{\rm Gyr}}
\newcommand{\dex}{\,{\rm dex}}
\newcommand{\Msun}{\,{\rm M}_\odot}
\newcommand{\kms}{\,{\rm km}\,{\rm s}^{-1}}
\newcommand{\onecolumn}{0.47\textwidth}
\newcommand{\figpath}[1]{./figures/#1}

\defcitealias{johnson_stellar_2021}{J21}
\defcitealias{leung_variational_2023}{L23}
\defcitealias{dubay_galactic_2024}{D24}

\shorttitle{Two-Infall Challenges}
\shortauthors{Dubay et al.}

\received{2025 July 31}
\revised{2026 February 13}
\accepted{2026 February 26}
\submitjournal{The Astrophysical Journal}

\begin{document}

\title{Challenges to the Two-Infall Scenario by Large Stellar Age Catalogs}

\author[0000-0003-3781-0747]{Liam O.\ Dubay}
\affiliation{\osuaffil}
\affiliation{\ccappaffil}
\email{dubay.11@osu.edu}
\author[0000-0001-7258-1834]{Jennifer A.\ Johnson}
\affiliation{\osuaffil}
\affiliation{\ccappaffil}
\email{}
\author[0000-0002-6534-8783]{James W.\ Johnson}
\affiliation{Observatories of the Carnegie Institution for Science, 813 Santa Barbara St., Pasadena CA 91101, USA}
\email{}
\author[0000-0002-2854-5796]{John D.\ Roberts}
\affiliation{\osuaffil}
\affiliation{\ccappaffil}
\email{}

\correspondingauthor{Liam O.\ Dubay}
\email{dubay.11@osu.edu}

\begin{abstract}
    Stars in the Milky Way disk exhibit a clear separation into two chemically distinct populations based on their [$\alpha$/Fe] ratios. This $\alpha$-bimodality is not a universal feature of simulated disk galaxies and may point to a unique evolutionary history. A popular and well-studied explanation is the two-infall scenario, which postulates that two periods of substantial accretion rates dominate the assembly history of the Galaxy. Thanks to recent advances in stellar age measurements, we can now compare this model to more direct measurements of the Galaxy's evolutionary timescales across the disk. We run multi-zone galactic chemical evolution models with a two-infall-driven star formation history and compare the results against abundance patterns from APOGEE DR17, supplemented with stellar ages estimated through multiple methods. Although the two-infall scenario offers a natural explanation for the $\alpha$-bimodality, it struggles to explain several features of the age--abundance structure in the disk. First, our models generically require a massive and long-lasting dilution event, but the data show that stellar metallicity is remarkably constant across much of the lifetime of the disk. This apparent age-independence places considerable restrictions upon the two-infall parameter space. Second, most local metal-rich stars in APOGEE have intermediate ages, yet our models predict these stars should either be very old or very young. Some of these issues can be mitigated, but not completely resolved, by pre-enriching the accreted gas to low metallicity. These restrictions also place limits on the role of merger events in shaping the chemical evolution of the thin disk.
\end{abstract}

\section{Introduction}
\label{sec:introduction}

\lettrine{G}{alactic chemical evolution} (GCE) studies aim to explain the metal abundance patterns observed in the Milky Way (MW) by modeling the star formation history and evolution of the Galaxy. A long-standing paradigm of GCE is that the metallicity of the interstellar medium (ISM) increases over time due to supernova enrichment from successive generations of stars \citep[e.g.,][]{tinsley_stellar_1979,matteucci_relative_1986}. In this view, one feature of the MW disk that is difficult to explain is the so-called ``$\alpha$-bimodality'': the presence of two populations of stars at similar metallicity but separated by their \aFe ratio \citep[e.g.,][]{bensby_exploring_2014}. The high-$\alpha$ sequence consists of old stars \citep[$\gtrsim9$ Gyr; e.g.,][]{pinsonneault_apokasc-3_2025} with super-Solar \aFe that are associated with the kinematic thick disk \citep[e.g.,][]{fuhrmann_nearby_1998}, while the low-$\alpha$ sequence is younger, has approximately Solar \aFe, and is associated with the kinematic thin disk. The $\alpha$-bimodality is present across the Galactic disk, but the relative strength of the high- and low-$\alpha$ sequences varies by location \citep{hayden_chemical_2015}. An $\alpha$-bimodality is not a universal feature in simulated MW-mass galaxies \citep[e.g.,][]{mackereth_origin_2018,parul_effect_2025}, and may not exist in M31 (\citealt{nidever_prevalence_2024}; but see also \citealt{kobayashi_fe_2023}), suggesting that it could point to a unique evolutionary history for the MW.

The two-infall scenario was first proposed by \citet{chiappini_chemical_1997} to explain the origin of the $\alpha$-bimodality in the MW. The basic premise is that the rate of gas accretion onto the disk occurs in two bursts. The relatively low infall rate between the two bursts causes star formation to slow down, allowing the gas metallicity to evolve between the high- and low-$\alpha$ sequences while producing few stars. \citet{chiappini_chemical_1997} successfully reproduced the Solar neighborhood abundance data available at the time, and subsequent studies refined the two-infall scenario to reproduce abundance data across the disk \citep[e.g.,][]{chiappini_abundance_2001,chiappini_oxygen_2003}, as well as the local surface densities of stars and gas and the local star formation and supernova rates \citep[e.g.,][]{romano_mass_2000,spitoni_galactic_2020,spitoni_remind_2024}. Others have explored the SN Ia delay-time distribution \citep{matteucci_effect_2009,palicio_analytic_2023}, galactic fountains \citep{spitoni_effects_2009}, radial gas flows \citep{spitoni_effects_2011,palla_chemical_2020}, variations in the star formation efficiency \citep{spitoni_effects_2011,palla_chemical_2020}, radial stellar migration \citep{spitoni_effect_2015,palla_mgfe_2022}, azimuthal abundance variations \citep{spitoni_2d_2019}, pre-enriched gas infall \citep{palla_chemical_2020,spitoni_remind_2024}, and additional episodes of gas accretion \citep{spitoni_beyond_2023,palla_mapping_2024}. Most recently, \citet{hegedus_reconstructing_2025} constrained the infall timescales using chemical abundances for $\sim400,000$ stars from the 19th data release of the Sloan Digital Sky Survey \citep[SDSS;][]{sdss_collaboration_nineteenth_2025}.

While previous studies have explored a large parameter space, most have assumed that the Galactic disk never had a substantial outflow. The presence of mass-loaded outflows in MW-like galaxies is heavily debated \citep[see discussion in][]{johnson_milky_2025}, but even if the MW is not currently ejecting a substantial outflow, it may have done so in the past. By neglecting Galactic outflows, previous studies of the two-infall scenario have been constrained in their choice of nucleosynthetic yields \citep{francois_evolution_2004} because of the yield--outflow degeneracy \citep[e.g.,][]{hartwick_chemical_1976,weinberg_equilibrium_2017,cooke_primordial_2022,johnson_dwarf_2023,sandford_strong_2024}. This degeneracy prohibits direct estimates of the yields from GCE models unless the effect of outflows is assumed to be insignificant. The predicted yields from CCSN models can vary substantially depending on the choice of initial mass function \citep{vincenzo_modern_2016} and the physics of black hole formation \citep{griffith_impact_2021}, yet few studies have investigated the effect of varying the yield scale on two-infall scenario predictions. Varying the yield scale while maintaining an evolutionary end point consistent with observations is straightforwardly achieved with simultaneous variations in the strength of outflows.

\begin{figure}
    \centering
    \includegraphics[width=\onecolumn]{\figpath{smooth_vs_twoinfall.pdf}}
    \caption{Chemical abundance tracks predicted by a two-infall model at $R_{\rm gal}=8\kpc$ (solid curve) versus a model with a smooth SFH at three different radii (dashed curves). The two-infall model adopts the fiducial parameters according to Table \ref{tab:parameters}, while the smooth SFH adopts the parameters of the ``inside-out'' model of \citet{johnson_stellar_2021}. Both models assume the \yZ{2} yield scale (see Table \ref{tab:yields}). The 2-D histogram shows the number of stars from APOGEE DR17 in the Solar neighborhood ($7\leq R_{\rm gal}<9\kpc$, $0\leq|z|<2\kpc$) in each bin of ([Fe/H], [O/Fe]).}
    \label{fig:smooth-vs-twoinfall}
\end{figure}

The two-infall scenario reproduces the full distribution of stellar abundances in the Solar neighborhood through a single, continuous evolutionary track. However, it is now understood that the stars that make up the Solar neighborhood are drawn from a wide range of birth radii \citep[e.g.,][]{sellwood_radial_2002,schonrich_chemical_2009,frankel_measuring_2018,lian_quantifying_2022,lehmann_probing_2024}. 
A number of GCE studies have reproduced the $\alpha$-bimdodality using a combination of radial migration and smooth, inside-out galaxy growth (e.g., \citealt{schonrich_chemical_2009,kubryk_evolution_2015,sharma_chemical_2021,chen_chemical_2023,prantzos_origin_2023}; but see also \citealt{johnson_stellar_2021,dubay_galactic_2024}). In this scenario, the local high-$\alpha$ population originates from the inner Galaxy, where the star formation rate peaked early in its history. Figure \ref{fig:smooth-vs-twoinfall} contrasts the chemical evolution track for the Solar neighborhood predicted by a two-infall model against tracks from the smooth SFH model of \citet{johnson_stellar_2021} at three different radii. 
If the effect of radial migration is significant, then the local $\alpha$-bimodality could be explained without dramatic changes to the local gas abundance over time.

As stellar age estimation techniques have improved over recent years, large catalogs have become available with age estimates for stars across a wide swath of the Galaxy. In a challenge to the traditional view of GCE, which expects the ISM metallicity to continually increase over time, these catalogs suggest that stellar metallicities are remarkably consistent with age across most of the thin disk \citep[e.g.,][]{spina_mapping_2022,magrini_gaia-eso_2023,willett_evolution_2023,carbajo-hijarrubia_occaso_2024,gallart_chronology_2024}. \citet{johnson_milky_2025} demonstrated that these results are readily explained by a scenario in which the Galaxy reaches a state of chemical equilibrium. In this so-called ``equilibrium scenario,'' the local metallicity is driven by the ratio of star formation to accretion at a given radius, which evolves to a constant over Gyr timescales. Whether the equilibrium metallicity is regulated by outflows, as in the  \citet{johnson_milky_2025} models, or by other factors such as a radial gas flow \citep{spitoni_effects_2011,bilitewski_radial_2012,sharda_interplay_2024}, the current data suggest that metal abundances in the ISM evolved minimally during the past $\sim8\Gyr$.

The availability of large stellar catalogs with [Fe/H], \aFe, {\it and} age covering a large portion of the Galactic disk enables much more comprehensive tests of the two-infall GCE scenario (and other models). Our study draws on the steadiness of the Milky Way stellar metallicity gradient found by \citet{johnson_milky_2025} and empirical constraints on the yield scale from \citet{weinberg_scale_2024}. We run multi-zone GCE models with a two-infall accretion history, radially-dependent mass-loaded outflows, and a prescription for radial migration tuned to a hydrodynamical simulation. We investigate the impact of the scale of SN yields and outflows, the strength of radial migration, the composition of the circumgalactic medium, and the local disk mass surface density ratio on GCE model predictions. We compare our results to abundance distributions across the disk from APOGEE DR17, and to age--abundance relations from multiple catalogs. We describe our observational sample in Section \ref{sec:observational-sample}, and we detail our chemical evolution models and parameter selection in Section \ref{sec:methods}. We compare our multi-zone model predictions to the data in Section \ref{sec:multizone-results}. We discuss possible extensions to our models in Section \ref{sec:discussion}, and we summarize our conclusions in Section \ref{sec:conclusions}.

\section{Observational Sample}
\label{sec:observational-sample}

\begin{table*}
    \centering
    \caption{Sample selection parameters from APOGEE DR17 (see Section \ref{sec:observational-sample}).}
    \label{tab:sample}
    \begin{tabular}{lll}
        \hline\hline
        \multicolumn{1}{c}{Parameter} & \multicolumn{1}{c}{Range or Value} & \multicolumn{1}{c}{Notes} \\
        \hline
        $\log g$            & $1.0 < \log g < 3.8$          & Select giants only \\
        $T_{\rm eff}$       & $3500 < T_{\rm eff} < 5500$ K & Reliable temperature range \\
        $S/N$               & $S/N > 80$                    & Required for accurate stellar parameters \\
        ASPCAPFLAG Bits     & $\notin$ 23                   & Remove stars flagged as bad \\
        EXTRATARG Bits      & $\notin$ 0, 1, 2, 3, or 4     & Select main red star sample only \\
        $R_{\rm gal}$     & $3 < R_{\rm gal} < 15\kpc$    & Eliminate bulge \& extreme outer-disk stars \\
        $|z|$               & $|z| < 2\kpc$                 & Eliminate halo stars \\
        \hline
        NN age error        & $\sigma_\tau/\tau < 40\%$     & Age uncertainty from \citet{leung_variational_2023} \\
        Uncertainty on [C/N]    & $\sigma_{\rm [C/N]}\leq0.1$     & Precise [C/N] ages from \citet{roberts_cn_2025} \\
        ${\rm [Fe/H]}$     & $\mathFeH\geq-0.4$            & Accurate [C/N] ages for upper RGB \& RC stars \\
        \hline
    \end{tabular}
\end{table*}

\begin{table}
    \centering
    \caption{Median and dispersion in APOGEE parameter uncertainties.}
    \label{tab:uncertainties}
    \begin{tabular}{l|cc}
\hline\hline
 & Median & Uncertainty  \\
Parameter & Uncertainty & Dispersion ($95\% - 5\%$) \\
\hline
${\rm [O/H]}$ & 0.019 & 0.031 \\
${\rm [Fe/H]}$ & 0.0089 & 0.006 \\
$\log_{10}(\tau_{\rm NN}/{\rm Gyr})$ & 0.1 & 0.16 \\
$\tau_{\rm [C/N]}/{\rm Gyr}$ & 1.6 & N/A \\
\hline
\end{tabular}

\end{table}

We compare our models against stellar abundances from the Apache Point Observatory Galactic Evolution Experiment \citep[APOGEE;][]{majewski_apache_2017} data release 17 \citep[DR17;][]{abdurrouf_seventeenth_2022}. APOGEE data were obtained from infrared spectrographs \citep{wilson_apache_2019} mounted on the 2.5-meter Sloan Foundation Telescope \citep{gunn_25_2006} at Apache Point Observatory and the Ir{\'e}n{\'e}e DuPont Telescope \citep{bowen_optical_1973} at Las Campanas Observatory. The data reduction pipeline is described by \citet{nidever_data_2015}, and the APOGEE Stellar Parameter and Chemical Abundance Pipeline (ASPCAP) is detailed by \citet{holtzman_abundances_2015}, \citet{garcia_perez_aspcap_2016}, and \citet{jonsson_apogee_2020}.

We obtain a sample of \num{171635} red giant branch and red clump stars with high-quality spectra using the selection criteria listed in Table \ref{tab:sample}, which are adapted from \citet{hayden_chemical_2015}. Table \ref{tab:uncertainties} presents the median statistical uncertainty and uncertainty dispersion ($95^{\rm th} - 5^{\rm th}$ percentile difference) of the calibrated [Fe/H] and [O/Fe]\footnote{
    In this paper, we use O as the representative $\alpha$-element in our GCE models and stellar abundance data. We refer to [O/Fe] and \aFe interchangeably, though observational studies often average several $\alpha$-element abundances to calculate \aFe.
} abundances for our sample. When calculating the galactocentric radius $R_{\rm gal}$ and midplane distance $z$ of each star, we use the \citet{bailer-jones_estimating_2021} photo-geometric distance estimates from {\it Gaia} Early Data Release 3 \citep{gaia_collaboration_gaia_2016,gaia_collaboration_gaia_2021} included in the APOGEE DR17 catalog and we adopt the Galactic coordinates of the Sun $(R,z)_\odot=(8.122, 0.0208)\kpc$ \citep{gravity_collaboration_detection_2018,bennett_vertical_2019}.

\subsection{Stellar Age Estimates}
\label{sec:age-estimates}

\begin{figure*}
    \centering
    \includegraphics[width=\textwidth]{\figpath{compare_age_catalogs.pdf}}
    \caption{Comparison between NN \citep{leung_variational_2023} and [C/N] \citep{roberts_cn_2025} ages for our sample. {\it Left:} Comparison of age estimates from both methods for all APOGEE stars. The solid black curve plots the rolling median (window size 1000 stars) of the [C/N] age estimate as a function of the NN age, the dashed line indicates the one-to-one correspondence, and the white error bars indicate the age uncertainty as a function of NN age. {\it Right:} The local ($7\leq R_{\rm gal}<9\kpc$, $|z|<0.5\kpc$) age--metallicity relation for each age estimation method. The black curve plots the rolling median [Fe/H] as a function of age.}
    \label{fig:compare-age-catalogs}
\end{figure*}

We supplement the APOGEE DR17 abundance data with two different age catalogs. The first is from \citet{leung_variational_2023}, who trained a variational encoder-decoder network on asteroseismic data for APOGEE red giants with $2.5<\log g<3.6$. Following the recommendations of \citet{leung_variational_2023}, we cut all stars which have an age uncertainty greater than 40\%. This produces a sample of \num{57607} stars with NN age estimates, of which \num{14871} are in the Solar neighborhood ($7\leq R_{\rm gal}<9\kpc$, $0\leq|z|<0.5\kpc$). The median uncertainty in $\log({\rm age})$ is 0.10 (see Table \ref{tab:uncertainties}), although the oldest stars typically have smaller uncertainties.

Our second age catalog utilizes the [C/N]--age relation calibrated by \citet{roberts_cn_2025} for red giant branch (RGB) and red clump (RC) stars. The relationship relies on the mass-dependent level of mixing during first dredge-up \citep[FDU;][]{iben_stellar_1967} to map the correlation of stellar mass, and hence age, with surface chemistry. This method has the benefit of providing age estimates for luminous giants ($\log g<2.5$), which increases the sample size at larger distances from the Sun. Uncertainties in the efficiency of FDU mixing and the RGB age--mass relationship mean that the ages are not trustworthy outside the range $1\sim10\Gyr$, but this is the age range most useful for our purposes anyway. Additional mixing effects in low-metallicity stars also prevent the relation from being applied to luminous RGB and RC stars with $\mathFeH<-0.4$; stars on the lower RGB do not suffer from this problem. 
We use the sample selection criteria from \citet{roberts_cn_2025}, and following their recommendation, we adopt a flat $1.64\Gyr$ age uncertainty for all stars.
With this relationship, we estimate ages for \num{124778} stars across the disk, including \num{21956} in the Solar neighborhood.

Figure \ref{fig:compare-age-catalogs} compares the NN ages from \citet{leung_variational_2023} and the [C/N]-derived ages from \citet{roberts_cn_2025} for our sample. Below an age of $\sim10\Gyr$, the two methods trace each other closely, although there is a small systematic offset for $\lesssim4\Gyr$ old stars. The Solar neighborhood age--metallicity relation (AMR) of \citet{leung_variational_2023} features substructure not seen in the \citet{roberts_cn_2025} AMR, possibly washed out due to the larger uncertainty in the latter sample. We compare these AMRs to others in the literature in Section \ref{sec:amr}. Overall, the AMRs derived from both methods are remarkably flat out to $\sim8-9\Gyr$, with variations of $\lesssim0.1\dex$ on the order of $1\Gyr$.

\section{Chemical Evolution Models \& Parameter Selection}
\label{sec:methods}

We run multi-zone GCE models using the Versatile Integrator for Chemical Evolution \citep[{\tt VICE};][]{johnson_impact_2020}. We wish to both approximate the Milky Way in gas and stellar density, present-day metallicity, and stellar abundance patterns, and to be able to tune parameters to help the two-infall scenario better match the observations and to reproduce the work of previous two-infall studies \citep[e.g.,][]{spitoni_galactic_2019,spitoni_apogee_2021,spitoni_remind_2024,palla_chemical_2020}.

The basic format of our models follows \citet{johnson_stellar_2021} and \citet{dubay_galactic_2024}. We set up a disk spanning $0\leq R_{\rm gal}<20\kpc$ that is divided into concentric rings of width $\delta R_{\rm gal}=100\,{\rm pc}$. We use a time-step of $\Delta t=10\,{\rm Myr}$, $n=8$ stellar populations per time-step per ring, and we run our models to a final time of $t_{\rm final}=13.2\,{\rm Gyr}$. We adopt a \citet{kroupa_variation_2001} initial mass function (IMF). Within each ring, chemical evolution proceeds according to a conventional one-zone GCE model with instantaneous mixing and continuous recycling. Stellar populations migrate between zones, allowing the long-lived progenitors of SNe Ia to enrich areas of the Galaxy outside of their birth zones, which couples the enrichment in nearby rings. We inhibit star formation past $R_{\rm gal}>15.5\kpc$, so stars in the outer 4.5 kpc of the model disk represent a purely migrated population.

We discuss our assumptions about the nucleosynthetic yields in Section \ref{sec:yields}, the outflow prescription in Section \ref{sec:outflows}, the star formation law in Section \ref{sec:sf-law}, the gas supply equation in Section \ref{sec:sfh}, the infall parameters in Section \ref{sec:infall-parameters}, the enrichment of the accreted gas in Section \ref{sec:cgm-enrichment}, and the stellar migration prescription in Section \ref{sec:migration}. We do not incorporate radial gas flows between the different zones, but we discuss their potential implications in Section \ref{sec:radial-flows}. Table \ref{tab:parameters} summarizes the model parameters and their fiducial values.

\begin{deluxetable*}{Cccl}
    \tablecaption{A summary of variables for our chemical evolution models (see discussion in Section \ref{sec:methods}).\label{tab:parameters}}
    \tablehead{
        \colhead{Quantity} & \colhead{Value or Range} & \colhead{Section} & \colhead{Description}
    }
    \startdata
        R_{\rm gal}         & $[0,20]\kpc$  & \ref{sec:methods}     & Galactocentric radius \\
        \delta R_{\rm gal}  & $0.1\kpc$     & \ref{sec:methods}     & Radial zone width \\
        R_{\rm SF}          & $15.5\kpc$    & \ref{sec:methods}     & Maximum radius of star formation \\
        t_{\rm final}       & $13.2\Gyr$    & \ref{sec:methods}     & Disk lifetime \\
        \Delta t            & $10\Myr$      & \ref{sec:methods}     & Time-step size \\
        n                   & 8             & \ref{sec:methods}     & Number of stellar populations formed per ring per timestep \\
        \dot M_r            & continuous    & \ref{sec:methods}     & Recycling rate \citep[][Equation 2]{johnson_impact_2020} \\
        {\rm IMF}           & \citet{kroupa_variation_2001} & \ref{sec:methods} & Initial stellar mass function \\
        y/Z_\odot           & $[1,2]$       & \ref{sec:yields}      & Scale of nucleosynthetic yields (see Table \ref{tab:yields}) \\
        N_{\rm Ia}/M_\star  & $[1.66,2.62]\times10^{-3}\Msun^{-1}$  & \ref{sec:yields}  & SNe Ia per unit mass of stars formed \\
        f_{\rm Ia}(t)       & Equations \ref{eq:plateau-dtd}, \ref{eq:powerlaw-dtd}     & \ref{sec:yields}  & Delay-time distribution of Type Ia supernovae \\
        t_{\rm Ia}          & $40\Myr$  & \ref{sec:yields}          & Minimum SN Ia delay time \\
        \eta_\odot          & $[0.2,1.4]$   & \ref{sec:outflows}    & Outflow mass-loading factor at $R_\odot$ ($\eta\equiv\dot\Sigma_{\rm out}/\dot\Sigma_\star$) \\
        R_\eta              & $5.0\kpc$     & \ref{sec:outflows}    & Exponential outflow scale radius \\
        \tau_\star          & Equation \ref{eq:sf-law}              & \ref{sec:sf-law}      & Star formation efficiency timescale ($\tau_\star\equiv\Sigma_g/\dot\Sigma_\star$) \\
        k                   & 1.5           & \ref{sec:sf-law}      & Star formation law exponent \citep{kennicutt_global_1998} \\
        M_{\rm \star,tot}   & $5.17\times 10^{10}\Msun$     & \ref{sec:sfh} & Total stellar mass of the disk \citep{licquia_improved_2015} \\
        f_\Sigma(R_\odot)   & $[0.12, 0.5]$     & \ref{sec:sfh}     & Local thick/thin disk surface density ratio \\
        R_1                 & $2.0\kpc$         & \ref{sec:sfh}     & Thick disk scale radius \\
        R_2                 & $2.5\kpc$         & \ref{sec:sfh}     & Thin disk scale radius \\
        f_{\rm in}(t|R_{\rm gal})   & Equation \ref{eq:infall-rate} & \ref{sec:sfh} & Time-dependence of the gas accretion rate \\
        \tau_1              & $1\Gyr$       & \ref{sec:infall-parameters}         & Timescale of the first infall epoch \\
        \tau_2(R_\odot)     & $15\Gyr$      & \ref{sec:infall-parameters}         & Timescale of the second infall epoch at the Solar annulus \\
        R_{\tau_2}          & $7\kpc$       & \ref{sec:sfh}         & Exponential scale radius of the second infall timescale \\
        t_{\rm max}         & $[2.2,3.2]\Gyr$     & \ref{sec:infall-parameters}         & Time of maximum gas infall (onset of second infall) \\
        \mathXH_{\rm CGM}   & $(-\infty,-0.5]$  & \ref{sec:cgm-enrichment}     & Metallicity of accreted gas \\
        \sigma_{\rm RM}(\tau,R_{\rm form})  & Equation \ref{eq:radial-migration}    & \ref{sec:migration} & Width of Gaussian controlling radial migration distance \\
        \sigma_{\rm RM8}    & $[2.68,5.0]\kpc$  & \ref{sec:migration}   & Radial migration strength (width of $\sigma_{\rm RM}$ for $\tau=8\Gyr$) \\
        z                   & $[-3,3]\kpc$      & \ref{sec:migration}   & Distance from Galactic midplane at present day \\
        h_z(\tau,R_{\rm final}) & Equation \ref{eq:scale-height}    & \ref{sec:migration} & Disk scale height \\
    \enddata
\end{deluxetable*}

\subsection{Nucleosynthetic Yields}
\label{sec:yields}

\begin{table}
    \centering
    \caption{Nucleosynthetic yields at each of the yield scales (see Section \ref{sec:yields}).}
    \begin{tabular}{c|cc}
\hline\hline
 & $y/Z_\odot=1$ & $y/Z_\odot=2$ \\
 & (empirical) & (theoretical) \\
\hline
$y_{\rm O}^{\rm CC}$ & \num{5.72e-03} & \num{1.14e-02} \\
$y_{\rm Fe}^{\rm CC}$ & \num{4.58e-04} & \num{9.15e-04} \\
$y_{\rm O}^{\rm Ia}$ & \num{0} & \num{0} \\
$y_{\rm Fe}^{\rm Ia}$ & \num{1.17e-03} & \num{1.83e-03} \\
\hline
$N_{\rm Ia}/M_\star\,[{\rm M}_\odot^{-1}]$ & \num{1.66e-03} & \num{2.62e-03} \\
\hline
\end{tabular}

    \label{tab:yields}
\end{table}

The population-averaged nucleosynthetic yields of CCSNe, $y_{\rm X}^{\rm CC}$, are uncertain to a degree that is significant for chemical evolution models. \citet{weinberg_scale_2024} used a measurement of the mean Fe yield of CC SNe by \citet{rodriguez_iron_2023} and the plateau in stellar \aFe abundances at low metallicity to infer population-averaged yields of $y/Z_\odot\approx1$.\footnote{
    In this work, we will use the $y/Z_\odot$ notation to refer to the scale set by the massive star $\alpha$-element yields; i.e., $y_{\alpha}^{\rm CC}/Z_{\alpha,\odot}$. We also clarify that these yields refer to the net metal production by stellar populations; the return of previously produced metals in the envelopes of dying stars is handled separately by {\tt VICE}.
} In other words, for every $1\Msun$ of stars formed, massive stars release a quantity of newly-synthesized $\alpha$-elements (e.g., O or Mg) equal to their mass fraction in the Sun. However, \citet{johnson_milky_2025} found that their GCE models with this yield scale approach present-day abundances too slowly to match the observed AMR. Previous multi-zone models using {\tt VICE} \citep[e.g.,][]{johnson_stellar_2021,dubay_galactic_2024} adopted higher yields ($y/Z_\odot\approx2.6$) based on \citet{chieffi_explosive_2004} and \citet{limongi_nucleosynthesis_2006}; however, in order to produce a realistic evolution of [O/Fe], those studies adopted a high integrated SN Ia rate compared to the measurement of \citet{maoz_star_2017}.

We therefore investigate yield sets at multiple scales. The CCSN yield of O is directly set by the Solar scale, $y_{\rm O}^{\rm CC}=(y/Z_\odot)Z_{\rm O,\odot}$\footnote{
    We adopt the \citet{asplund_chemical_2009} Solar abundances: $Z_{\rm O,\odot}=5.72\times10^{-3}$ and $Z_{\rm Fe,\odot}=1.29\times10^{-3}$. We note that for this empirical scaling, choosing different Solar abundances \citep[e.g.,][]{magg_observational_2022} would lead us to change the yields proportionally, so that our GCE model predictions in Solar-scaled abundances would actually be unchanged.
}, because we assume that all O is produced by CCSNe. For Fe, the CCSN yield is set by the \aFe plateau at low metallicity, $\mathOFe_{\rm CC}$, such that $y_{\rm Fe}^{\rm CC}=(y/Z_\odot) Z_{\rm Fe,\odot} 10^{-\mathOFe_{\rm CC}}$ \citep[for further discussion on the empirical yield scale and the CCSN plateau, see][]{weinberg_scale_2024}. We set the plateau at $\mathOFe_{\rm CC}=+0.45$, which corresponds to an Fe yield from CCSNe of $y_{\rm Fe}^{\rm CC}=4.58\times10^{-4}(y/Z_\odot)$. Our yield sets are presented in Table \ref{tab:yields}. We consider \yZ{1} representative of the empirical yield scale and \yZ{2} representative of theoretical predictions.

The SN Ia yield of Fe, $y_{\rm Fe}^{\rm Ia}$, is calibrated so that our models reach $\mathOFe\approx0.0$ by $t=13.2\,{\rm Gyr}$. The fifth row of Table \ref{tab:yields} reports the integrated SN Ia rate or total number of SNe Ia per unit mass of star formation,
\begin{equation}
    \frac{N_{\rm Ia}}{M_\star} = \frac{y_{\rm Fe}^{\rm Ia}}{\bar m_{\rm Fe}^{\rm Ia}},
    \label{eq:snia-rate}
\end{equation}
for each yield set, assuming a mean Fe yield per SN Ia of $\overline m_{\rm Fe}^{\rm Ia}=0.7$ M$_\odot$ \citep{mazzali_common_2007,howell_effect_2009}. The rate for the \yZ{1} yield set is slightly higher than the volumetric rate of $N_{\rm Ia}/M_\star=(1.3\pm0.1)\times10^{-3}\,{\rm M}_\odot^{-1}$ reported by \citet{maoz_star_2017}, but is consistent with their measurement of $N_{\rm Ia}/M_\star=(1.6\pm0.3)\times10^{-3}\,{\rm M}_\odot^{-1}$ for field galaxies. The rate for the \yZ{2} yield set is consistent with the measurement of $N_{\rm Ia}/M_\star=(2.2\pm1.0)\times10^{-3}\,{\rm M}_\odot^{-1}$ by \citet{maoz_type-ia_2012}.

Unlike CCSNe, SNe Ia populate a broad distribution of delay times between progenitor formation and explosion. The time-dependent SN Ia rate per unit mass of star formation is defined as
\begin{equation}
    R_{\rm Ia}(t) = 
    \begin{cases}
        \frac{N_{\rm Ia}}{M_\star}
        \frac{f_{\rm Ia}(t)}{\int_{t_{\rm Ia}}^{t_{\rm final}} f_{\rm Ia}(t') dt'}, & t \ge t_{\rm Ia} \\
        0 & t < t_{\rm Ia},
    \end{cases}
    \label{eq:dtd-function}
\end{equation}
where $t_{\rm Ia}=40\,{\rm Myr}$ is the minimum SN Ia delay time and $f_{\rm Ia}(t)$ is the un-normalized form of the DTD. Motivated by recent results suggesting that a large fraction of long-delayed SNe Ia improves agreement with the Milky Way's high-$\alpha$ sequence \citep{palicio_analytic_2023,dubay_galactic_2024}, we adopt a wide plateau DTD of the form
\begin{equation}
    \label{eq:plateau-dtd}
    f_{\rm Ia}(t) =
    \begin{cases}
        1, & t < 1\,{\rm Gyr} \\
        (t/1\,{\rm Gyr})^{-1.1}, & t \ge 1\,{\rm Gyr}.
    \end{cases}
\end{equation}
We also explore a simple power-law DTD,
\begin{equation}
    f_{\rm Ia}^{\rm plaw}(t) = (t/1\,\rm{Gyr})^{-1.1}.
    \label{eq:powerlaw-dtd}
\end{equation}
Equation \ref{eq:powerlaw-dtd} is consistent with SN Ia surveys \citep[e.g.,][]{maoz_star_2017} and results in more prompt Fe production.
We discuss the implications of using the power-law DTD in Section \ref{sec:abundance-distributions}.

Many previous two-infall studies have adopted the yields of \citet{francois_evolution_2004}, who in turn adapted those of \citet{woosley_evolution_1995} for CCSNe and \citet{iwamoto_nucleosynthesis_1999} for SNe Ia to provide a better fit between GCE models and local abundance data. Notably, the yields for O and Fe were left unchanged from the original studies. However, because \citet{woosley_evolution_1995} report gross yields without detailed initial abundances for their CCSN progenitors, and because \citet{francois_evolution_2004} do not provide population-averaged yields, it is difficult to make a comparison with our yields. Ultimately, \citet{francois_evolution_2004} report that their GCE models are insensitive to changes in the CCSN yield of O by a factor of 2, so we consider it reasonable to explore both yield scales presented in Table \ref{tab:yields}.

\subsection{Outflows}
\label{sec:outflows}

Mass-loaded outflows are a useful tool for scaling the endpoint of GCE models. \citet{weinberg_equilibrium_2017} showed that in the case of exponentially declining star formation, the O abundance approaches an equilibrium at
\begin{equation}
    \label{eq:equilibrium}
    Z_{\rm O,eq} = \frac{y_{\rm O}^{\rm CC}}{1 + \eta - r - \tau_\star/\tau_{\rm SFH}},
\end{equation}
where $r=0.4$ is the instantaneous recycling parameter for a \citet{kroupa_variation_2001} IMF, $\tau_\star$ is the star formation efficiency timescale, $\tau_{\rm SFH}$ is the $e$-folding timescale of the star formation history, and $\eta\equiv \dot\Sigma_{\rm out}/\dot\Sigma_\star$ is the outflow mass-loading factor. Motivated by Equation \ref{eq:equilibrium}, we adopt a different outflow mass-loading factor at the Solar radius $\eta_\odot\equiv\eta(R=R_\odot)$ for each of the yield sets in Table \ref{tab:yields}. 
Measurements of gas-phase \citep[e.g.,][]{mendez-delgado_gradients_2022} and stellar (Figure \ref{fig:yield-outflow}) abundances indicate that the Solar neighborhood ISM is presently close to Solar metallicity, so we adjust $\eta_\odot$ to ensure that $Z_{\rm O,eq}\approx Z_{\rm O,\odot}$.
Table \ref{tab:models} reports the value of $\eta_\odot$ in each of our models.

Variations in $\eta$ with $R_{\rm gal}$ produce variations in $Z_{\rm O,eq}$. To produce a radial [O/H] gradient, we adopt a prescription for the outflow mass-loading factor from \citet{johnson_milky_2025} of
\begin{equation}
    \eta(R_{\rm gal}) = \eta_\odot \exp\Big(\frac{R_{\rm gal}-R_\odot}{R_\eta}\Big),
\end{equation}
where $R_\eta$ is the exponential outflow scale radius and $R_\odot=8\kpc$. We adopt $R_\eta=5$ kpc, a lower value than in \citet{johnson_milky_2025}, so that our $y/Z_\odot=1$ model produces a radial abundance gradient of $\nabla\mathOH_{\rm eq}\approx-0.06\,{\rm dex}\kpc^{-1}$, consistent with recent measurements from HII regions \citep{mendez-delgado_gradients_2022} and stars \citep{myers_open_2022,johnson_milky_2025}. We note, however, that mass-loaded outflows are not a necessary ingredient for the results of this study (see discussion in Section \ref{sec:radial-flows}). A one-zone model with the fiducial parameters, $\eta=0$, and \yZ{0.8}, predicts a similar abundance evolution and nearly identical stellar abundance distributions to the fiducial model with $\eta=0.2$ and \yZ{1}.

\subsection{The Star Formation Law}
\label{sec:sf-law}

We use a power-law star formation prescription of $\dot\Sigma_\star\propto\Sigma_g^k$, with $k=1.5$ following \citet{kennicutt_global_1998}. Previous work with this GCE model \citep[e.g.,][]{johnson_stellar_2021,dubay_galactic_2024} assumed a three-component power-law, but we adopt a single power-law to allow for a more direct comparison with previous two-infall studies \citep[e.g.,][]{spitoni_remind_2024}. In detail, we calculate the star formation efficiency (SFE) timescale
\begin{equation}
    \label{eq:sf-law}
    \tau_\star \equiv \frac{\Sigma_g}{\dot\Sigma_\star} = 
    \begin{cases}
        \varepsilon(t) \tau_{\rm mol}(t),   & \Sigma_g \ge \Sigma_{g,0} \\
        \varepsilon(t) \tau_{\rm mol}(t) \Big(\frac{\Sigma_g}{\Sigma_{g,0}}\Big)^{-1/2}, & \Sigma_g < \Sigma_{g,0},
    \end{cases}
\end{equation}
with $\Sigma_{g,0} = 10^8\,{\rm M}_\odot\kpc^{-2}$ and $\tau_{\rm mol}(t)=\tau_{\rm mol,0}(t/t_0)^\gamma$, where $\gamma=1/2$, $t_0=13.8\,{\rm Gyr}$ and $\tau_{\rm mol,0}=2\,{\rm Gyr}$ are observationally calibrated \citep{leroy_star_2008}. Previous two-infall studies \citep[e.g.,][]{spitoni_galactic_2019,spitoni_galactic_2020,palla_chemical_2020} have adopted a higher SFE during the first infall epoch than the second, which we emulate through a pre-factor
\begin{equation}
    \label{eq:sfe-prefactor}
    \varepsilon(t) = 
    \begin{cases}
        0.5, & t < t_{\rm max} \\
        1.0, & t \ge t_{\rm max}.
    \end{cases}
\end{equation}
A lower value of $\varepsilon(t<t_{\rm max})$ leads to more efficient star formation, and therefore more rapid enrichment, during the first infall epoch. However, the pre-factor has virtually no effect on the overall [O/Fe] distribution because the model is normalized to produce the same thick-to-thin-disk mass ratio regardless of the details of the star formation law. We find $\epsilon(t<t_{\rm max})=0.2$ or $1.0$ produce similar results in one-zone models.

\subsection{The Gas Supply}
\label{sec:sfh}

We run {\tt VICE} in ``infall mode,'' where we specify the gas infall density $\dot\Sigma_{\rm in}$ and the SFE timescale $\tau_\star$ as functions of time at each radius. The gas surface density $\Sigma_g$ and star formation rate $\dot\Sigma_\star$ are calculated from these two inputs as a natural outcome of the time-stepping solution, assuming zero initial gas mass in all zones.

The infall rate as a function of time and galactocentric radius is generically described by
\begin{equation}
    \label{eq:infall-rate}
    \dot\Sigma_{\rm in}(t,R_{\rm gal}) = A f_{\rm in}(t|R_{\rm gal}) g(R_{\rm gal}),
\end{equation}
where $g(R_{\rm gal})=\Sigma_\star(R_{\rm gal}) / \Sigma_\star(R_{\rm gal}=0)$ is the stellar density gradient, $f_{\rm in}$ is the infall rate over time, and $A$ is a normalization constant. Because we incorporate mass-loaded outflows, $A$ is not analytically solvable, so we numerically integrate $\dot\Sigma_\star(t,R_{\rm gal})$ and then normalize $\dot\Sigma_{\rm in}$ to produce a total disk stellar mass of $M_{\rm \star,tot}=(5.17\pm1.11)\times 10^{10}\,{\rm M}_\odot$ \citep{licquia_improved_2015} and match the stellar surface density gradient of \citet{bland-hawthorn_galaxy_2016}.

The infall rate is described by two successive decaying exponential components, which form the thick and thin disks. The un-normalized form of the infall rate is
\begin{multline}
    \label{eq:twoinfall-ifr}
    f_{\rm in}(t|R_{\rm gal}) \propto e^{-t/\tau_1} + \\ H(t-t_{\rm max}) f_{2/1} (R_{\rm gal}) \exp\Big(\frac{-(t-t_{\rm max})}{\tau_2(R_{\rm gal})}\Big),
\end{multline}
where $\tau_1$ and $\tau_2(R_{\rm gal})$ are the first and second infall timescales, respectively, $t_{\rm max}$ is the onset of the second infall, $f_{2/1}$ is the ratio of the second infall amplitude to the first, and $H$ is the Heaviside step function,
\begin{equation*}
    H(x) \equiv 
    \begin{cases}
        1, & x \ge 0 \\
        0, & x < 0.
    \end{cases}
\end{equation*}
We vary $\tau_2$ exponentially with $R_{\rm gal}$ and adopt a scale radius $R_{\tau_2}=7\kpc$ to match the stellar age gradients in MW-like spirals observed by \citet{sanchez_spatially_2020}. We numerically calculate $f_{2/1}$ for each zone such that the resulting stellar disk has a surface density ratio of the thick and thin disks of
\begin{equation}
    f_\Sigma(R) \equiv \frac{\Sigma_1(R)}{\Sigma_2(R)} = f_\Sigma(R_\odot) e^{(R-R_\odot)\cdot(1/R_2 - 1/R_1)}.
\end{equation}
We adopt a thick disk scale radius of $R_1=2.0\kpc$, a thin disk scale radius of $R_2=2.5\kpc$, and a fiducial value for the local surface density ratio of $f_\Sigma(R_\odot)=0.12$ \citep{bland-hawthorn_galaxy_2016}. 

The thick-to-thin disk density ratio is especially important for our GCE models as it controls the quantity of gas accreted during each infall epoch. Our fiducial value of $f_\Sigma(R_\odot)=0.12$ is on the low end of literature estimates, which range from $f_\Sigma(R_\odot)\sim0.06-0.6$ \citep[e.g.,][]{gilmore_new_1983,siegel_star_2002,juric_milky_2008,mackereth_age-metallicity_2017,fuhrmann_local_2017}. Previous two-infall studies have adopted a similarly broad range of values (e.g., $f_\Sigma(R_\odot)=0.18$ from \citealt{spitoni_apogee_2021}; $f_\Sigma(R_\odot)=0.4$ from \citealt{spitoni_remind_2024}). We therefore explore values up to $f_\Sigma(R_\odot)=0.5$ in our multi-zone models in Section \ref{sec:multizone-results}.

\subsection{Infall Rate Parameter Selection}
\label{sec:infall-parameters}

Previous studies have adopted a wide range of parameter values for Equation \ref{eq:twoinfall-ifr} \citep[see, e.g.,][]{nissen_high-precision_2020,spitoni_galactic_2020,spitoni_apogee_2021,hegedus_reconstructing_2025}. After exploring the effect of the infall timescales in one-zone models, we adopt $\tau_1=0.3\Gyr$, $\tau_2(R_\odot)=15\Gyr$, and $t_{\rm max}=3.2\Gyr$ (i.e., a lookback time of $10\Gyr$; see Appendix \ref{app:infall-parameters} for more details). \citet{spitoni_galactic_2019} discuss the effect of these parameters on the GCE model predictions in detail. We also adopt $t_{\rm max}=2.2\Gyr$ for some models to closely match the evolution of the \citet{palicio_analytic_2023} analytic models (see Figure \ref{fig:yield-outflow}).

Our value of $t_{\rm max}$ is inconsistent with the median stellar asteroseismic age of the thick disk, $9.14\pm0.05\Gyr$ \citep{pinsonneault_apokasc-3_2025}, although other methods recover older ages \citep[e.g.,][]{sahlholdt_characterizing_2022}. We have run models with $t_{\rm max}=4.2\Gyr$ and found that it worsens agreement with APOGEE data in the [O/Fe]--[Fe/H] plane. Additionally, such a late second infall would exacerbate the dilution challenges discussed in Section \ref{sec:age-abundance}.

\begin{figure}
    \centering
    \includegraphics[width=\onecolumn]{\figpath{star_formation_history.pdf}}
    \caption{The evolution of (a) the infall surface density, (b) the star formation surface density, (c) the gas surface density, and (d) the star formation efficiency timescale as a function of time for our fiducial multi-zone model with $y/Z_\odot=1$. Each panel plots the evolution in six different zones of width $\delta R_{\rm gal}=0.1\kpc$, color-coded by Galactocentric radius.}
    \label{fig:sfh}
\end{figure}

Figure \ref{fig:sfh} plots the star formation history for several different zones from our fiducial model with $y/Z_\odot=1$. In the inner Galaxy, the infall rate is highest at $t=0$, and the star formation rate peaks at $t\approx6\,{\rm Gyr}$. In the outer Galaxy, the infall rate at $t_{\rm max}$ is similar to $t=0$, and the star formation rate is highest at the present day. The SFE timescale spikes near $t=0$ and $t_{\rm max}$, but otherwise increases throughout the model's duration, reaching a present-day value of $\tau_\star\approx2\,{\rm Gyr}$ in the inner disk and $\tau_\star\approx 9\,{\rm Gyr}$ in the outer disk.

\subsection{Infalling Gas Metallicity}
\label{sec:cgm-enrichment}

In most of our models, we assume the infalling gas is pristine (i.e., $Z_{\rm in}=0$). However, previous GCE studies suggest that some level of enrichment of the infalling gas can improve agreement with observations \citep[e.g.,][]{palla_chemical_2020,spitoni_remind_2024,johnson_milky_2025}. The circumgalactic medium (CGM) from which the infalling gas is drawn could be previously enriched, possibly from contributions from Galactic outflows, gas stripped from dwarf galaxies, or from SNe in the halo. The Milky Way's CGM is diffuse, multiphase, and inhomogeneous, making it difficult to study \citep[e.g.,][]{tumlinson_circumgalactic_2017,mathur_probing_2022}; still, observations have confirmed the existence of metals at non-Solar abundance ratios in the CGM \citep[e.g.,][]{das_discovery_2019,das_hot_2021,gupta_supervirial_2021}. We investigate models where the infalling gas is pre-enriched to a metallicity
\begin{equation}
    \label{eq:pre-enrichment}
    Z_{\rm in}(t) = (1 - e^{-t/\tau_{\rm rise}}) Z_\odot 10^{\mathXH_{\rm CGM}}.
\end{equation}
In this case, the metallicity rises from 0 with a timescale $\tau_{\rm rise}=2\,{\rm Gyr}$ and plateaus at $\mathFeH_{\rm CGM}$. We assume the Solar-scaled abundance of the accreted gas is the same for all elements (i.e., $\mathXH_{\rm CGM}=\mathOH_{\rm CGM}$).

\subsection{Stellar Migration}
\label{sec:migration}

This study is not the first to apply a prescription for radial migration to a two-infall GCE model. \citet{spitoni_effect_2015} explored the effect of migration speeds of order $\sim1\,{\rm km}{\rm s}^{-1}$ on the metallicity distribution of the Solar neighborhood. 
Our implementation, based on the functional form of \citet{frankel_measuring_2018}, allows stellar populations to migrate between zones during each timestep, which allows us to explore the effects of radial migration over the entire disk. \citet{palla_mgfe_2022} compared the \citet{spitoni_effect_2015} prescription to that of \citet{frankel_measuring_2018} and found a similar effect on stellar abundance distributions.

The distance a stellar population born at $R_{\rm form}$ migrates over its age $\tau$ is drawn from a Gaussian centered at 0 with standard deviation
\begin{equation}
    \sigma_{\rm RM}(\tau,R_{\rm form}) = \sigma_{\rm RM8} \Big(\frac{\tau}{8\,{\rm Gyr}}\Big)^{0.33} \Big(\frac{R_{\rm form}}{8\kpc}\Big)^{0.61},
    \label{eq:radial-migration}
\end{equation}
where we adopt $\sigma_{\rm RM8}=2.68\kpc$ as the fiducial value for the strength of radial migration from \citet{dubay_galactic_2024}. This is smaller than the value of $\sigma_{\rm RM8}=3.6\kpc$ found by \citet{frankel_measuring_2018}, but in Section \ref{sec:age-abundance} we explore the effect of a stronger migration prescription.

All stellar populations are born at the Galactic midplane and are assigned a final midplane distance $z$ drawn from the distribution
\begin{equation}
    p(z|\tau,R_{\rm final}) = \frac{1}{4 h_z} {\rm sech}^2\Big(\frac{z}{2 h_z}\Big),
    \label{eq:sech-pdf}
\end{equation}
\citep{spitzer_dynamics_1942} where $R_{\rm final}$ is the final Galactocentric radius of the stellar population. The scale height $h_z$ is given by
\begin{equation}
    h_z(\tau,R_{\rm final}) = \Big(\frac{0.24\kpc}{e^2}\Big) \exp\Big(\frac{\tau}{7\,{\rm Gyr}} + \frac{R_{\rm final}}{6\kpc}\Big).
    \label{eq:scale-height}
\end{equation}
The final midplane distance is assigned at the end of the model run and therefore does not affect the chemical evolution. The parameters of Equations \ref{eq:radial-migration} and \ref{eq:scale-height} were chosen to fit the stellar migration patterns in the {\tt h277} hydrodynamical simulation \citep{christensen_implementing_2012}. A more complete discussion of the migration scheme and its consequences can be found in Appendix C of \citet{dubay_galactic_2024}.

An important distinction between our method and that of \citet{spitoni_effect_2015} is that SNe Ia from long-lived progenitors contribute Fe to each zone they migrate through, not just their birth zone. This is important because the median delay time of our SN Ia DTD is $\sim2$ Gyr, for which the width of the migration distribution is $\sigma_{\rm RM}\approx2$ kpc (Equation \ref{eq:radial-migration}). Therefore, a significant fraction of SN Ia progenitors born in a given zone will enrich a different region of the Galaxy.

\section{Multi-Zone Model Results}
\label{sec:multizone-results}

We run 11 multi-zone models at the \yZ{1} and \yZ{2} yield scales. We vary the SN Ia DTD, the migration strength $\sigma_{\rm RM8}$, the thick-to-thin disk mass ratio $f_\Sigma(R_\odot)$, and the enrichment of the infalling gas ${\rm [X/H]}_{\rm CGM}$. Our models and relevant parameters are summarized in Table \ref{tab:models}. In the remainder of this Section, we present results from these models and compare against APOGEE data and stellar age estimates.

\begin{deluxetable*}{lCCCcDDC}
    \tablecaption{Summary of our multi-zone models and relevant parameters.\label{tab:models}}
    \tablehead{
        \colhead{Name} & \colhead{$y/Z_\odot$} & \colhead{$\eta_\odot$} & \colhead{$t_{\rm max}$ [Gyr]} & \colhead{SN Ia DTD} & \multicolumn2c{$\sigma_{\rm RM8}$ [kpc]} & \multicolumn2c{$f_\Sigma(R_\odot)$} & \colhead{${\rm [X/H]}_{\rm CGM}$}
    }
    \decimals
    \startdata
        {\tt yZ1-fiducial}  & 1 & 0.2   & 3.2 & Equation \ref{eq:plateau-dtd}  & 2.68  & 0.12  & $-\infty$   \\
        {\tt yZ1-migration} & 1 & 0.2   & 3.2 & Equation \ref{eq:plateau-dtd}  & 5.0   & 0.12  & $-\infty$   \\
        {\tt yZ1-diskratio} & 1 & 0.2   & 3.2 & Equation \ref{eq:plateau-dtd}  & 2.68  & 0.50  & $-\infty$   \\
        {\tt yZ1-preenrich} & 1 & 0.6   & 3.2 & Equation \ref{eq:plateau-dtd}  & 2.68  & 0.12  & -0.5      \\
        {\tt yZ1-best}      & 1 & 0.4   & 3.2 & Equation \ref{eq:plateau-dtd}  & 3.6   & 0.25  & -0.7      \\
        \hline
        {\tt yZ2-fiducial}  & 2 & 1.4   & 3.2 & Equation \ref{eq:plateau-dtd}  & 2.68  & 0.12  & $-\infty$   \\
        {\tt yZ2-earlyonset}& 2 & 1.4   & 2.2 & Equation \ref{eq:plateau-dtd}  & 2.68  & 0.12  & $-\infty$   \\
        {\tt yZ2-powerlaw}  & 2 & 1.4   & 3.2 & Equation \ref{eq:powerlaw-dtd} & 2.68  & 0.12  & $-\infty$   \\
        {\tt yZ2-diskratio} & 2 & 1.4   & 3.2 & Equation \ref{eq:plateau-dtd}  & 2.68  & 0.50  & $-\infty$   \\
        {\tt yZ2-preenrich} & 2 & 2.4   & 3.2 & Equation \ref{eq:plateau-dtd}  & 2.68  & 0.12  & -0.5      \\
        {\tt yZ2-best}      & 2 & 1.8   & 2.2 & Equation \ref{eq:plateau-dtd}  & 3.6   & 0.25  & -0.7      \\
    \enddata
\end{deluxetable*}

\subsection{Dilution \& Re-enrichment}
\label{sec:age-abundance}

\begin{figure}
    \centering
    \includegraphics[width=\onecolumn]{\figpath{gas_abundance_evolution.pdf}}
    \caption{The ISM abundance evolution at $R_{\rm gal}=8\kpc$ of three multi-zone models at different yield scales (see Table \ref{tab:models}), each of which feature a major dilution event $10-11\Gyr$ ago. The 2-D histograms show the APOGEE stellar age--abundance distributions in the Solar neighborhood ($7\leq R_{\rm gal}<9\kpc$, $|z|<0.5\kpc$), adopting the \citet{leung_variational_2023} ages. The black curve (points) plots the rolling median (binned mode) of the abundance data as a function of age, and the gray error bars along the bottom of each panel indicate the median age and abundance errors as a function of age. The gray dashed curve plots the abundance evolution of the \citet{palicio_analytic_2023} analytic model, which our {\tt yZ2-earlyonset} model (green curve) closely matches. The left-hand marginal panels show the predicted (blue, pink, and green) and observed (gray) stellar abundance distributions, which are boxcar-smoothed with a width of 0.05 dex for visual clarity.}
    \label{fig:yield-outflow}
\end{figure}

Figure \ref{fig:yield-outflow} shows the APOGEE age--abundance relations in the Solar annulus, using the NN ages from \citet{leung_variational_2023}. While the age--abundance distributions are clumpy in places (see Section \ref{sec:amr}), we caution that the APOGEE selection function can affect the stellar age distribution \citep[see][]{imig_galactic_2025}. As shown by \citet{johnson_milky_2025}, the peak of the MDF is less susceptible to modification by radial migration, making it a more reliable proxy for ISM chemistry at a given lookback time than the mean or median. Strikingly, we observe that ${\rm mode}(\mathOH)\gtrsim0$ over the past 10 Gyr.

We over-plot the ISM abundance evolution in our \yZ{1} and \yZ{2} models with the fiducial parameters at $R_{\rm gal}=8\kpc$ for comparison. The models are in broad agreement with each other and with the data at lookback times of $\lesssim5\Gyr$. At ages of $\sim5-9\Gyr$, the models underpredict the APOGEE metallicities by up to half an order of magnitude, a natural consequence of the dilution associated with the second infall event. This discrepancy is modestly reduced in the \yZ{2} model, which re-enriches more rapidly. However, this models exacerbates tensions with the age--[O/Fe] relation because the O abundance increases more before enrichment from second-infall SNe Ia becomes important, making the $\alpha$-enhancement from the ensuing starburst more pronounced. Overall, none of our models provide a good match to these age trends. Figure \ref{fig:yield-outflow} illustrates a point that will arise as a theme throughout the remainder of this paper, which is that the signatures of a substantial dilution event and subsequent re-enrichment, characteristic of the two-infall scenario, are simply not present in the observed age--abundance trends.

Figure \ref{fig:yield-outflow} also plots the {\tt yZ2-earlyonset} model, for which we adopt $t_{\rm max}=2.2\Gyr$ to more closely match the evolution of the \citet{palicio_analytic_2023} analytic model. For reference, we also plot the analytic solution using the Chemical Evolution Analytic Package \citep[ChEAP;][]{palicio_analytic_2023}. The magnitude of the dilution at $t_{\rm max}$ is $\sim0.2$ dex smaller than for the {\tt yZ2-fiducial} model, and the gas abundance is more consistent with the $5-9\Gyr$ old stars. However, the minimum stellar age for the thick disk is $1\Gyr$ older, worsening the tension with the measurement of \citet{pinsonneault_apokasc-3_2025}.

\begin{figure*}
    \centering
    \includegraphics[width=\linewidth]{\figpath{stellar_abundance_evolution.pdf}}
    \caption{Stellar age--abundance relations predicted by multi-zone models at the \yZ{1} yield scale (see Table \ref{tab:yields}). Each point represents a stellar population drawn from the Solar neighborhood near the midplane ($7\leq R_{\rm gal}< 9\kpc$, $|z| < 0.5\kpc$) color-coded by its birth radius. In this and subsequent figures, a Gaussian scatter is applied to each point according to the median age and abundance uncertainties in Table \ref{tab:uncertainties}. For visual clarity, we plot only a random mass-weighted sample of \num{10000} points in each panel. The dashed curve plots the predicted ISM abundance at $R_{\rm gal}=8\kpc$, and the solid black curve plots the rolling median stellar abundance. The red curve plots the rolling median abundance of the APOGEE sample, and the shaded regions are the 16th--84th percentile ranges. Each column shows results from a different multi-zone model: {\bf (a)} our fiducial model, {\tt yZ1-fiducial}; {\bf (b)} a model with stronger radial migration, {\tt yZ1-migration}; {\bf (c)} a model with a higher local thick-to-thin disk ratio, {\tt yZ1-diskratio}; and {\bf (d)} a model with pre-enriched gas infall, {\tt yZ1-preenrich} (see Table \ref{tab:models} for details).}
    \label{fig:abundance-evolution-params}
\end{figure*}

Chemical evolution models that assume the \yZ{1} (empirical) yield scale already struggle to match the local AMR even with a smooth SFH \citep[see also][]{johnson_milky_2025}. The problem is exacerbated in the two-infall case because of the delayed dilution event---the approach to equilibrium is ``reset'' by the second infall. The dilution of the ISM then gets baked into the stellar abundance record.
Figure \ref{fig:abundance-evolution-params} shows stellar age--abundance relations predicted by models with \yZ{1}. The model with the fiducial parameters (column (a); {\tt yZ1-fiducial}) shows similar discrepancies with the \citet{leung_variational_2023} NN age--[O/H] relation as in Figure \ref{fig:yield-outflow}. The evolution of [Fe/H] is similar, but the approach to the final metallicity is slower because of the additional delay imposed by SNe Ia. We smoothed simulated data points based on the median age and abundance errors in Table \ref{tab:uncertainties} to incorporate these errors in the model. Despite the effects of radial migration and statistical errors, the stellar populations mostly scatter around the gas phase abundance track, at least for $\tau\lesssim9\Gyr$. Between ages $\sim5-9\Gyr$, the predicted median stellar metallicity is $\sim0.5\dex$ lower than observed, well below the 16\textsuperscript{th} percentile of the data.

We next attempt to mitigate the dilution and late-time evolution problems for the \yZ{1} yield scale. First, the observed rise in the median metallicity of stars in the $5-9\Gyr$ age range could be due to radial migration, as those stars were probably not born in-situ, but rather migrated from the metal-rich inner Galaxy \citep{feuillet_age-resolved_2018}. Column (b) of Figure \ref{fig:abundance-evolution-params} presents model {\tt yZ1-migration}, which has a stronger migration prescription of $\sigma_{\rm RM8}=5\kpc$. As a result, the stars that make up the present-day Solar neighborhood are drawn from a wider range of birth $R_{\rm gal}$, producing a broader abundance distribution at fixed age. Even though this prescription is extreme compared to our fiducial model or the measurement of \citet{frankel_measuring_2018}, the model still significantly under-predicts the median metallicity of $\sim5-9\,{\rm Gyr}$ old stars.

Next, we explore model {\tt yZ1-diskratio}, which has a local thick-to-thin disk surface density ratio $f_\Sigma(R_\odot)=0.5$, $\sim4$ times larger that the fiducial value. This is higher than most of the constraints from population counts or GCE models (see Section \ref{sec:sfh}). Column (c) of Figure \ref{fig:abundance-evolution-params} shows that requiring a more massive thick disk can reduce the dilution and recent evolution of the ISM because more of the gas disk is built up during the first infall phase. Model {\tt yZ1-diskratio} produces the best agreement of the four models in Figure \ref{fig:abundance-evolution-params} to the observed age--[Fe/H] relation (second row). However, agreement with the observed age--[O/Fe] relation is poor. The model predicts a much flatter trend than observed, under-predicting the median [O/Fe] by $\sim0.1$ dex in the $\sim5-9\Gyr$ age range.

Finally, we investigate model {\tt yZ1-preenrich}, where the infalling gas is enriched before accreting onto the disk. Column (d) of Figure \ref{fig:abundance-evolution-params} shows results for the case where ${\rm [X/H]}_{\rm CGM}=-0.5$, the highest metallicity allowed by the local low-$\alpha$ population. 
Pre-enriched infall at this level mitigates but does not completely solve the two discrepancies. The dilution effect of the second infall is reduced to the $\sim0.3$-dex level as the gas which replenishes the Galaxy's reservoir is no longer pristine; however, the width of the stellar abundance distribution at any given age is also reduced, since the accreting gas is less chemically different from the ISM, diminishing the effects of dilution. This model also narrows the [O/Fe] distribution of mono-age populations (almost all the model stars fall within the $1\sigma$ band of the data), which could be compensated for by stronger radial migration.

Overall, no modification to the \yZ{1} model is able to completely overcome the issues that dilution and re-enrichment naturally face when confronted with a flat AMR. Pre-enrichment of the accreted gas and a higher disk mass ratio can reduce the discrepancy with the data, but these options introduce new issues in the age--[O/Fe] plane. 

\subsection{Abundance Evolution Across the Disk}
\label{sec:disk-evolution}

\begin{figure*}
    \centering
    \includegraphics[width=\textwidth]{\figpath{mdf_evolution.pdf}}
    \caption{Decomposition of the present-day MDF by stellar age across the Galactic disk. 
    The gray curve plots the total present-day MDF in each region. The distributions in all panels are restricted to $|z|<0.5\kpc$ and boxcar-smoothed with a width of {0.1 dex} for visual clarity. Rows (a) and (b) present the distributions from the {\tt yZ1-fiducial} and {\tt yZ2-fiducial} models, respectively (see Table \ref{tab:models}). Row (c) presents the distributions from APOGEE with ages derived from [C/N] abundances by \citet{roberts_cn_2025} (see Section \ref{sec:age-estimates}). In row (c), the vertical blue dotted lines mark the mode of the distribution of the $1-2\kpc$ age bin for reference, and the gray dashed line marks the cut at $\mathFeH\ge-0.4$ for upper RGB and RC stars.}
    \label{fig:mdf-evolution}
\end{figure*}

The discrepancies between the predicted and observed abundance evolution in the Solar neighborhood discussed in Section \ref{sec:age-abundance} persist across the Galactic disk. Figure \ref{fig:mdf-evolution} breaks down the MDF by stellar age across five radial bins for the {\tt yZ1-fiducial} and {\tt yZ2-fiducial} models. For the APOGEE sample, we use the [C/N]-derived age estimates due to the larger sample size in the most distant regions of the disk; we limit the comparison to ages in the range $1-10$ Gyr because the [C/N] ages are most reliable in this range, as discussed in Section \ref{sec:age-estimates}.

The predictions of both models in Figure \ref{fig:mdf-evolution} show a clear trend in [Fe/H] with age at all radii. The MDF shifts consistently toward high metallicity when moving from older to younger stars. The distance between the $1-2\Gyr$ and $2-4\Gyr$ age bins is smaller for the \yZ{2} model because of the faster approach to equilibrium (see also Figure \ref{fig:yield-outflow}). In the Solar annulus (center column), the peak of the MDF of $6-8\Gyr$ old stars is $0.3\dex$ lower in the \yZ{2} model than observed, and $0.4\dex$ lower in the \yZ{1} model.

In contrast, the APOGEE data show remarkably little evolution in [Fe/H] up to ages of $\sim8\Gyr$ at all radii. Row (c) of Figure \ref{fig:mdf-evolution} shows that the MDF broadens with age, but its peak remains constant across this range. The mode [Fe/H] for the youngest stars (indicated by the vertical blue dotted line) is nearly the same as for the $6-8\Gyr$ old stars. At $R_{\rm gal}<7\kpc$, the MDF skews to lower [Fe/H] more noticeably with age, but its mode does not shift by more than $\sim0.1\dex$. It is difficult to draw conclusions about the outer Galaxy because the mode [Fe/H] is close to the metallicity cut at $\mathFeH>-0.4$ for luminous giants (represented by the vertical gray dashed line), which comprise the majority of stars in the sample at that distance. The remarkable consistency of the MDF over time, the result that motivated the equilibrium scenario proposed by \citet{johnson_milky_2025}, contrasts with the predictions of our fiducial models.

\subsection{The [O/Fe] Distribution}
\label{sec:abundance-distributions}

\begin{figure}
    \centering
    \includegraphics[width=\onecolumn]{\figpath{ofe_feh_density.pdf}}
    \caption{The density of stars in the [O/Fe]--[Fe/H] plane predicted by multi-zone models with (a) $y/Z_\odot=1$ and (b) $y/Z_\odot=2$. The curves plot the ISM abundance at the Solar annulus over time, and the alternating black and white segments mark time intervals of 1~Gyr. The gray curves in the marginal panels plot the APOGEE stellar abundance distributions.
    Stars in both panels are restricted to the region defined by $7\leq R_{\rm gal}< 9\kpc$ and $|z|<2\kpc$, and the model output has been re-sampled to match the APOGEE stellar $|z|$ distribution.}
    \label{fig:ofe-feh-density}
\end{figure}

In the two-infall scenario, the low-$\alpha$ loop governs the chemical evolution of the thin disk \citep[see Section 6 of][]{spitoni_galactic_2019}. However, careful inspection of the marginal [O/Fe] distributions in Figure \ref{fig:yield-outflow} reveals a different morphology: the models predict {\it three} peaks in the [O/Fe] distribution, whereas the data show only two. The location of the intermediate peak varies depending on the yields and model parameters, but is always present. This morphology remains essentially consistent in our multi-zone models as well, despite the inclusion of radial mixing and vertical dispersion of stars.

Figure \ref{fig:ofe-feh-density} illustrates the origin of the intermediate-$\alpha$ peak predicted by the two-infall model. Both the models with $y/Z_\odot=1$ and $y/Z_\odot=2$ predict an over-density of stars near the abundance turn-over ($\mathFeH\approx-0.4$, $\mathOFe\approx0.1-0.2$), which is not seen in the APOGEE sample. The local maximum in $dN_\star/d\mathOFe$ comprises stars which formed in the second infall before SN Ia enrichment began to drive down [O/Fe]. This prediction should therefore arise in {\it any} two-infall model regardless of its specific parameters, but its impact can be mitigated through parameter choices that act to compress the distance between the low- and intermediate-$\alpha$ peaks, as in the $y/Z_\odot=1$ model in Figure \ref{fig:yield-outflow}.

\begin{figure*}
    \centering
    \includegraphics[width=\textwidth]{\figpath{ofe_distributions.pdf}}
    \caption{Normalized stellar [O/Fe] distributions predicted by multi-zone models with \yZ{2} (a--d) and as observed by APOGEE (e). Each row presents stellar distributions within a range of absolute midplane distance $|z|$ reported on the far right, and the vertical scale is consistent across each row. Within each panel, the distributions are color-coded according to the bin in galactocentric radius $R_{\rm gal}$ from which they are drawn. For visual clarity, each distribution is smoothed with a box-car of width 0.05 dex.
    Each column shows the distributions predicted from a different multi-zone model: {\bf (a)} the fiducial model with \yZ{2}, {\tt yZ2-fiducial}; {\bf (b)} a model with a power-law DTD, {\tt yZ2-powerlaw}; {\bf (c)} a model with a higher local thick-to-thin disk ratio, {\tt yZ2-diskratio}; and {\bf (d)} a model with pre-enriched gas infall, {\tt yZ2-preenrich} (see Table \ref{tab:models} for details).}
    \label{fig:ofe-distribution-parameters}
\end{figure*}

Figure \ref{fig:ofe-distribution-parameters} compares [O/Fe] distributions from across the Galactic disk predicted by models with \yZ{2}. We present the distributions in multiple bins of $|z|$ as well as $R_{\rm gal}$ because the observed pattern varies as a function of midplane distance, and because the APOGEE selection function over-emphasizes high-$|z|$, and therefore high-$\alpha$, stars in the full sample \citep[see Figure 5 from][]{vincenzo_distribution_2021}. The {\tt yZ2-fiducial} model\footnote{While not shown, the {\tt yZ2-earlyonset} model predicts nearly identical stellar [O/Fe] distributions to {\tt yZ2-fiducial}.} (column (a)) predicts a high density of stars at $\mathOFe\approx+0.2$, where the data instead show a trough. The peak is most pronounced in the inner disk ($R_{\rm gal}=3-5\kpc$) because of the shorter infall timescale.

We attempt to mitigate the intermediate peak through several parameter choices. First, we substitute our fiducial SN Ia DTD with a simple power-law (Equation \ref{eq:powerlaw-dtd}; {\tt yZ2-powerlaw}),
which reduces the median SN Ia delay time from $\sim2\,{\rm Gyr}$ to $\sim0.5\,{\rm Gyr}$. As shown in column (b), this has the intended effect on the low-$\alpha$ sequence, but it also entirely eliminates the high-$\alpha$ peak. \citet{dubay_galactic_2024} discuss in detail why such a DTD is disfavored by Milky Way stellar abundances for several different SFHs, including for the two-infall scenario. 

Next, in model {\tt yZ2-diskratio} (column (c)) we increase the local thick-to-thin disk surface density ratio by a factor of 4, to $f_\Sigma(R_\odot)=0.5$. The result is a true bimodal abundance distribution, with a more prominent high-$\alpha$ peak than in the other models. However, the location of the high-$\alpha$ (low-$\alpha$) peak is $\sim0.1$ dex higher (lower) than observed.

Finally, in model {\tt yZ2-preenrich} (column (d)) the metallicity of the infalling gas increases to $\mathXH_{\rm CGM}=-0.5$ at late times. This model predicts similar [O/Fe] distributions to the $y/Z_\odot=1$ case. We assume that the infalling gas has $\mathOFe_{\rm CGM}=0$ at all times; an alternate model with $\mathOFe_{\rm CGM}=+0.3$ shifted the distribution towards higher [O/Fe], worsening agreement with observations. 

In summary, either an enhanced disk mass ratio or pre-enriched infall can improve agreement with the observed thin disk abundances for the \yZ{2} case. These parameters also help the model better fit the AMR, as shown in Section \ref{sec:age-abundance} for the \yZ{1} case.

\subsection{The Best Model}
\label{sec:ofe-feh-best}

\begin{figure*}
    \centering
    \includegraphics[width=\linewidth]{\figpath{ofe_feh_best.pdf}}
    \caption{Stellar abundance distributions across the disk predicted by the {\tt yZ2-best} model (see Table \ref{tab:models}). Each panel presents a random mass-weighted sample of \num{10000} stellar populations color-coded by age. The solid (dashed) contours enclose 30\% (80\%) of the APOGEE stars in each region.}
    \label{fig:ofe-feh-best}
\end{figure*}

Motivated by the results of the previous sections, we attempt to construct a model that solves all of the issues that have been outlined thus far. Our ``best attempt'' model {\tt yZ2-best} uses the \yZ{2} yield set to flatten the local AMR (Figure \ref{fig:yield-outflow}), $t_{\rm max}=2.2\Gyr$ to better match the local [O/Fe]--[Fe/H] distribution (Section \ref{sec:infall-parameters}), pre-enriched infall at the level of $\mathXH_{\rm CGM}=-0.7$ to reduce the dilution at $t_{\rm max}$ (Figure \ref{fig:abundance-evolution-params}), slightly stronger outflows with $\eta_\odot=1.8$ to maintain the local equilibrium at Solar metallicity, moderately stronger radial migration with $\sigma_{\rm RM8}=3.6\kpc$ to widen the local metallicity dispersion (Figure \ref{fig:abundance-evolution-params}), and a greater local disk ratio $f_\Sigma(R_\odot)=0.25$ to reduce the width of the low-$\alpha$ distribution and beef up the high-$\alpha$ sequence (Figure \ref{fig:ofe-distribution-parameters}). Our choices for $\mathXH_{\rm CGM}$, $\sigma_{\rm RM8}$, and $f_\Sigma(R_\odot)$ are more moderate than the illustrative examples in the previous section. Our focus is on qualitative rather than quantitative agreement with the data, and thus we do not attempt to find the optimal set of parameters.

Figure \ref{fig:ofe-feh-best} shows the stellar [O/Fe]--[Fe/H] distributions in different ranges of $R_{\rm gal}$ and $|z|$ color-coded by age as predicted by the {\tt yZ2-best} model. This model is generally successful at reproducing the observed abundance distributions, especially in the inner Galaxy and close to the midplane. However, due to the stronger migration prescription and higher thick-to-thin disk ratio, implemented to solve the dilution problem in the Solar neighborhood, the high-$\alpha$ sequence is too prominent in the outer Galaxy. In general, the predicted distributions do not align with the data as well at large midplane distances ($1\leq|z|<2\kpc$), but this may reflect inaccuracies in our prescription for vertical heating (see Section \ref{sec:migration}).

\subsection{Local Age Patterns}

\begin{figure*}
    \centering
    \includegraphics[width=\textwidth]{\figpath{lmr_ages.pdf}}
    \caption{{\it Top:} The median stellar age as a function of [O/Fe] and [Fe/H] in the Solar annulus ($7\leq R_{\rm gal}<9\kpc$, $0\leq|z|<2\kpc$). The left and center panels show the {\tt yZ1-best} and {\tt yZ2-best} models, with ${\rm [X/H]}_{\rm CGM}=-0.7$, $f_\Sigma(R_\odot)=0.25$, and $\sigma_{\rm RM8}=3.6\kpc$ (see Table \ref{tab:models}). The model output has been re-sampled to match the APOGEE stellar $|z|$ distribution, and a Gaussian scatter has been applied to the abundances and ages according to Table \ref{tab:uncertainties}. The right panel plots the results from APOGEE using the \citet{leung_variational_2023} NN age catalog. The contours show the density of stars in the [Fe/H]--[O/Fe] plane, and the vertical dashed line denotes the boundary for locally metal-rich (LMR) stars.
    {\it Bottom}: Stellar age distributions in the Solar annulus for all stars (black) and LMR stars (gray). The left and center panels plot the mass-weighted age distributions predicted by the models, and the right panel plots the NN ages for APOGEE stars.}
    \label{fig:lmr-ages}
\end{figure*}

The two-infall scenario makes a fundamental prediction about the local stellar age distribution: the stars born at the tail end of the thick and thin disk epochs are adjacent to each other in abundance space, meaning the two-infall scenario predicts a bimodal age distribution for metal-rich stars. This prediction is apparent in any of the panels in Figure \ref{fig:ofe-feh-best}, especially where $|z|<0.5\kpc$. The top row of panels in Figure \ref{fig:lmr-ages} presents the median stellar age as a function of [O/Fe] and [Fe/H] predicted by the {\tt yZ1-best} and {\tt yZ2-best} models (see Table \ref{tab:models}) and observed in APOGEE using the \citet{leung_variational_2023} NN ages. While the models predict a fairly accurate distribution of stars in abundance space, especially for the low-$\alpha$ population, the stellar age patterns are starkly different. In both models, there is a sharp divide in the median stellar age when moving from the thick disk ($\tau\ge9\Gyr$) to the thin disk ($\tau\lesssim5\Gyr$). In contrast, the data show a smoother age gradient between the two populations.

We further highlight the discrepant age patterns in the bottom panels of Figure \ref{fig:lmr-ages}, which compare the overall stellar age distribution against that of the locally metal rich (LMR) stars, defined here as $\mathFeH\ge+0.1$.\footnote{
    While the amplitude of the peaks is sensitive to the precise location of the LMR cutoff, the dearth of intermediate-age stars is insensitive to adjustments of $\pm0.05\dex$.
} For APOGEE, the distributions of all stars and only LMR stars are similar, both peaking near $\sim5\Gyr$, although very few of the LMR stars have ages $\gtrsim10\Gyr$. Our two-infall models predict an overall age distribution that is similar to the data, but both models predict a distinctly different age distribution for LMR stars. For the \yZ{1} model, the LMR age distribution is strongly bimodal; the two populations reflect the metal-rich endpoints of the two successive infall epochs. For the \yZ{2} model, the LMR ages are not bimodal but are heavily skewed to younger ($\lesssim6\Gyr$) ages (note that the models have different values of $t_{\rm max}$, which affects the predicted age distributions). Both models predict few LMR stars with intermediate ($\sim4-8\Gyr$) ages, whereas these stars dominate the LMR population in APOGEE. A dearth of intermediate-age, high-metallicity stars is a natural outcome of a substantial dilution event, which is a defining feature of the two-infall scenario. None of the adjustments that we explored in Sections \ref{sec:age-abundance} or \ref{sec:abundance-distributions} substantially increase the proportion of intermediate-age stars.

Other discrepancies between the models and data in Figure \ref{fig:lmr-ages} reflect our specific parameter choices, not intrinsic limitations of the two-infall framework. First, the predicted ages of high-$\alpha$ stars are $\sim2$ Gyr older than observed in APOGEE. Delaying the onset of thick disk formation (e.g., by introducing an early ``simmering'' phase of low star formation efficiency proposed by \citealt{conroy_birth_2022}) or delaying $t_{\rm max}$ could alleviate this tension, but it would not improve the dearth of intermediate-aged LMR stars. Second, some of the youngest APOGEE stars have [O/Fe] ratios $\sim0.05-0.1\dex$ lower than predicted by the models (see also the bottom panel of Figure \ref{fig:yield-outflow}). Increasing $y_{\rm Fe}^{\rm Ia}$ by a factor of $10^{0.05}\approx1.1$ brings the present-day gas abundance in line with the data but worsens agreement with the overall stellar [O/Fe] distribution. Finally, there is a population of $\sim8-10\Gyr$ old stars at sub-Solar [O/Fe], formed during the period of rapid dilution immediately after $t_{\rm max}$. These stars form a small percentage of the overall distribution \citep[Figure \ref{fig:ofe-feh-best}; see also Figure 11 from][]{spitoni_remind_2024} but they occupy a unique part of abundance space. Adopting a longer $\tau_1$ or smaller $t_{\rm max}$ could shift this population to higher [O/Fe], where it would be obscured by the rest of the low-$\alpha$ sequence (see Appendix \ref{app:infall-parameters}).

\section{Discussion}
\label{sec:discussion}

In this work, we have highlighted a few predictions of the two-infall scenario and compared against trends from large stellar age catalogs. We have explored a wide range of parameter space, but our investigation is not intended to be exhaustive. In this section, we discuss further extensions to the two-infall model as well as other recent evidence of the Galaxy's evolution.

\subsection{The Age--Metallicity Relation}
\label{sec:amr}

A few recent studies report the discovery of multiple disconnected sequences in the local AMR \citep[e.g.,][]{sahlholdt_characterizing_2022,xiang_time-resolved_2022,anders_spectroscopic_2023}. \citet{nissen_high-precision_2020} argued that the existence of multiple sequences is evidence for the two-infall scenario. However, \citet{plotnikova_chemical_2024} showed that this pattern can arise due to selection effects, and the APOGEE selection function in particular affects the observed stellar age distribution \citep{imig_galactic_2025}. Furthermore, \citet{chen_recent_2025} demonstrated that multiple sequences in the local AMR can arise naturally due to radial migration, even in the absence of major accretion or starburst events. While the model of \citet{chen_recent_2025} does not reproduce the MW's abundance patterns, most notably lacking a strong $\alpha$-bimodality, it demonstrates that the two-infall scenario is not the only explanation for multiple sequences in the local AMR.

Figure \ref{fig:yield-outflow} shows hints of multiple sequences in the \citet{leung_variational_2023} AMR, specifically where an under-density of $\sim5\Gyr$ old stars coincides with a $\sim0.2\dex$ drop in [O/H] and [Fe/H]. This is less pronounced, but otherwise similar to, the sequences found by \citet{anders_spectroscopic_2023}. We note a systematic offset between the two catalogs around the same age \citep[see Figure 6 of][]{anders_spectroscopic_2023}. On the other hand, the [C/N] AMR of \citet{roberts_cn_2025} is flatter and does not appear to have multiple sequences (Figure \ref{fig:compare-age-catalogs}), although the larger age uncertainties may wash out such patterns.

Even if these patterns in the AMR are real, the two-infall scenario does not offer a natural explanation. The difference in metallicity between the sequences in our sample is $\sim0.2\dex$, smaller than the dilution predicted by any of our models. Additionally, the transition between the sequences occurs $\sim5\Gyr$ ago \citep[matching][]{anders_spectroscopic_2023}, much younger than the oldest thin disk stars \citep{pinsonneault_apokasc-3_2025}. Interestingly, the transition does align with the first pericentric passage of the Sagittarius dwarf galaxy, which has been linked to periods of enhanced star formation in the disk \citep{ruiz-lara_recurrent_2020}. It is possible that the Milky Way has seen multiple episodes of high star formation without experiencing the dilution associated with the two-infall scenario.

\subsection{Third Accretion Episode}

Motivated by evidence of a recent period of enhanced star formation across the MW disk \citep{ruiz-lara_recurrent_2020}, \citet{spitoni_beyond_2023} extended the two-infall scenario with a recent $(\lesssim3\Gyr)$ third accretion episode. In contrast to the two-infall model of \citet{spitoni_apogee_2021}, which predicted a present-day gas metallicity of ${\rm [M/H]}\approx+0.3$ in the Solar neighborhood, \citet{spitoni_beyond_2023} argued that the gas dilution resulting from the third infall could explain a population of young stars with slightly sub-Solar abundances discovered in {\it Gaia} DR3 \citep{recio-blanco_gaia_2023}.

Similarly, \citet{palla_mapping_2024} invoked a late-time accretion episode to explain largely age-independent metallicites in open clusters across the disk. Open clusters trace younger populations, and most of their clusters are $\lesssim2\Gyr$ old. A third infall can help explain trends within this limited age range, but when compared to the full range of stellar age data, it faces the same challenges as the second infall. Successive epochs of dilution and re-enrichment will always struggle to explain chemical abundances that remain nearly constant with age over a $\sim10\Gyr$ period. As discussed in Section \ref{sec:age-abundance}, pre-enriched infall could reduce the magnitude of the dilution from a recent accretion episode, but the stellar age data place tight constraints on the parameter space.

\subsection{Outflows Versus Radial Gas Flows}
\label{sec:radial-flows}

The role of Galactic winds in the Milky Way's evolution is uncertain. Some simulations of CCSN feedback predict that ejected material falls back onto the disk on relatively short timescales \citep{spitoni_galactic_2008,spitoni_effects_2009} close to its point of origin \citep{melioli_hydrodynamical_2008,melioli_hydrodynamical_2009}, suggesting that the effect on GCE should be minimal. However, the effects of feedback in simulations are sensitive to its implementation \citep[e.g.,][]{li_effects_2020,hu_code_2023}; other simulations of MW-like galaxies with different prescriptions do produce mass-loaded outflows \citep[e.g.,][]{brook_hierarchical_2011,gutcke_nihao_2017,nelson_first_2019,peschken_angular_2021,kopenhafer_seeking_2023}. Empirically, mass-loading is observed in nearby starburst galaxies \citep[e.g.,][]{lopez_temperature_2020,cameron_duvet_2021,lopez_x-ray_2023} but not MW-like systems because the predicted column densities are below current detection limits \citep[see reviews by][]{veilleux_cool_2020,thompson_theory_2024}. Additionally, observed wind velocities in external galaxies suggest that outflowing gas may remain gravitationally bound \citep{concas_two-faces_2019}.

Radial gas flows are a potential alternative to the outflow prescription that we use in this paper. Within the two-infall paradigm, \citet{spitoni_effects_2011} found that inward flow velocities of a few $\kms$ can steepen the predicted radial metallicity gradient to match observations. Similarly, \citet{palla_chemical_2020,palla_mapping_2024} used an inward flow with a speed of $\sim1\kms$ to reproduce the observed radial metallicity profile in the absence of Galactic outflows. Recently, \citet{johnson_constraints_2025} considered several possible scenarios for the processes that drive radial gas flows and confirmed that radial gas flows can help to regulate the overall metal abundance and radial gradient.

In our models, the outflows regulate the overall metallicity in a radially dependent manner, which \citet{johnson_milky_2025} demonstrated can improve agreement with age trends across the Galactic disk. The outflow is simpler in implementation than a radial gas flow; we can simply adjust the mass loading factors up or down in combination with a given scale of yields. Additionally, our models with pre-enriched CGM are a step toward modeling the re-accretion of material from Galactic winds, as suggested by simulations of Galactic fountains and the observations of \citet{concas_two-faces_2019}. Outflows allow us to easily consider models that predict metallicity to evolve on different timescales but reach similar abundances at the present day (see, e.g., Figure \ref{fig:yield-outflow}). We have therefore focused on models with outflows in this paper for the sake of simplicity, but constant radial gas flow models should have similar effects.

\subsection{Star Formation Hiatus Models}
\label{sec:sfe-hiatus}

The two-infall scenario produces an $\alpha$-bimodality due to a few Gyr long quiescent period between the two major accretion epochs. However, significant gas accretion is not a necessary ingredient to produce a hiatus in star formation. Investigating a simulated galaxy from the Illustris TNG50 suite \citep{pillepich_first_2019,nelson_first_2019,nelson_illustristng_2019}, \citet{beane_rising_2024} argued that its MW-like $\alpha$-bimodality is the result of a brief ($\sim300\,{\rm Myr}$) quiescent period caused by the formation of a bar. In this scenario, the newly-formed bar funnels gas into the galaxy's central black hole, prompting a burst of AGN activity which suppresses star formation. Alternatively, \citet{johnson_thats_2025} showed that a retrograde merger can induce much of the ISM to sink to the center of the Galaxy, lowering the local gas surface density and thereby creating an $\alpha$-bimodality. The low mass ratio of the merger ($\sim3:1$) means the thin disk is not formed directly from gas accreted during the merger, but rather that radial gas flows induced by the merging galaxy's trajectory alter the stellar \aFe distributions. Both of these scenarios explain the observed \aFe distribution without invoking significant dilution and re-enrichment of the ISM.

\section{Summary \& Conclusions}
\label{sec:conclusions}

We have compared the predictions of the two-infall scenario against abundance data from APOGEE DR17 supplemented with age estimates using two different methods. We ran multi-zone GCE models at two different yield scales with prescriptions for radially-dependent outflows and stellar migration. While the two-infall scenario can explain the distinct high- and low-$\alpha$ sequences, it faces challenges in matching the age--abundance structure of the full disk. We explored multiple parameter modifications to bring the model predictions closer to the data, including the yield scale, radial migration strength, metallicity of the accreted gas, thick-to-thin disk mass ratio, and the SN Ia DTD. We found that our two-infall models faced the following challenges:

\begin{itemize}
    \item By the model's design, the Galaxy undergoes a period of rapid dilution and subsequent re-enrichment. Models at a lower yield scale and weaker outflows predict slower re-enrichment. By contrast, observed stellar metallicities are roughly constant over the past $\sim10\Gyr$. Pre-enriching the accreted gas can reduce the magnitude of the dilution, but cannot eliminate it entirely.
    \item Due to this re-enrichment, our models predict that the MDF shifts to higher metallicity with decreasing age throughout the disk. This contrasts with the APOGEE data, which show little change in the mode of the MDF over the past $\sim6-8$ Gyr.
    \item For metal-rich stars, the two-infall model predicts a sharp divide in the stellar age distribution between the thick and thin disk populations: local metal-rich stars should either be very old or very young. In contrast, the data show a broad, unimodal age distribution, with most of the metal-rich stars having intermediate ages. This discrepancy could not be reconciled using any of the model variations we explored.
    \item The ``turn-over'' in the evolution of [O/Fe] following the second infall produces a double-peaked low-$\alpha$ sequence with a fundamentally different abundance structure than observed, especially for models with higher yields. A low yield set (\yZ{1}) coupled with lower outflows, or pre-enrichment of the infalling gas, can bring the stellar [O/Fe] distributions more in line with the data. The parameter space is nonetheless restricted by the need to suppress this feature.
\end{itemize}

We have not found a GCE model, two-infall or otherwise, that is capable of reproducing all of the observations. The ``inside-out'' models of \citet{johnson_stellar_2021} are better able to predict a slowly evolving metallicity gradient, but they do not predict an $\alpha$-bimodality. Even a smoothly-evolving SFH struggles to reproduce the flat AMR at the lower empirical yield scale \citep{johnson_milky_2025}. Therefore, we do not rule out the two-infall scenario in favor of other models. We emphasize, though, that by its very construction, the two-infall scenario trades success at reproducing certain observations for challenges at others.

The apparent age-independence of stellar abundances in the disk places considerable restrictions upon the two-infall parameter space because it predicts a substantial dilution event at the start of the thin disk epoch. If the equilibrium scenario of \citet{johnson_milky_2025} is accurate, then it restricts the two-infall scenario more than other GCE models. Restrictions on the two-infall scenario also translate more broadly to other merger-dominated SFHs, which must explain the MW's $\alpha$-bimodality without significantly diluting the metallicity of the ISM.

\section*{Acknowledgements}

We are grateful to Prof.\ David Weinberg for his thoughtful comments and input on this paper, and to Prof.\ Christopher Kochanek for his useful comments on the manuscript. We also thank Dr.\ Emanuele Spitoni for useful conversations about this work. We thank the anonymous reviewer for thoughtful and constructive comments on the manuscript. Finally, we thank the attendees of OSU's Galaxy Hour for many useful discussions over the course of this project.

LOD and JAJ acknowledge support from National Science Foundation grant no.\ AST-2307621. JAJ and JWJ acknowledge support from National Science Foundation grant no.\ AST-1909841.
LOD acknowledges financial support from an Ohio State University Fellowship.
JWJ acknowledges financial support from a Carnegie Theoretical Astrophysics Center postdoctoral fellowship.

Funding for the Sloan Digital Sky 
Survey IV has been provided by the 
Alfred P.\ Sloan Foundation, the U.S.\ 
Department of Energy Office of 
Science, and the Participating 
Institutions. 

SDSS-IV acknowledges support and 
resources from the Center for High 
Performance Computing  at the 
University of Utah. The SDSS 
website is \url{www.sdss4.org}.

SDSS-IV is managed by the 
Astrophysical Research Consortium 
for the Participating Institutions 
of the SDSS Collaboration including 
the Brazilian Participation Group, 
the Carnegie Institution for Science, 
Carnegie Mellon University, Center for 
Astrophysics | Harvard \& 
Smithsonian, the Chilean Participation 
Group, the French Participation Group, 
Instituto de Astrof\'isica de 
Canarias, The Johns Hopkins 
University, Kavli Institute for the 
Physics and Mathematics of the 
Universe (IPMU) / University of 
Tokyo, the Korean Participation Group, 
Lawrence Berkeley National Laboratory, 
Leibniz Institut f\"ur Astrophysik 
Potsdam (AIP),  Max-Planck-Institut 
f\"ur Astronomie (MPIA Heidelberg), 
Max-Planck-Institut f\"ur 
Astrophysik (MPA Garching), 
Max-Planck-Institut f\"ur 
Extraterrestrische Physik (MPE), 
National Astronomical Observatories of 
China, New Mexico State University, 
New York University, University of 
Notre Dame, Observat\'ario 
Nacional / MCTI, The Ohio State 
University, Pennsylvania State 
University, Shanghai 
Astronomical Observatory, United 
Kingdom Participation Group, 
Universidad Nacional Aut\'onoma 
de M\'exico, University of Arizona, 
University of Colorado Boulder, 
University of Oxford, University of 
Portsmouth, University of Utah, 
University of Virginia, University 
of Washington, University of 
Wisconsin, Vanderbilt University, 
and Yale University.

This work has made use of data from the European Space Agency (ESA) mission
{\it Gaia} (\url{https://www.cosmos.esa.int/gaia}), processed by the {\it Gaia}
Data Processing and Analysis Consortium (DPAC,
\url{https://www.cosmos.esa.int/web/gaia/dpac/consortium}). Funding for the DPAC
has been provided by national institutions, in particular the institutions
participating in the {\it Gaia} Multilateral Agreement.

We would like to acknowledge the land that The Ohio State University occupies is the ancestral and contemporary territory of the Shawnee, Potawatomi, Delaware, Miami, Peoria, Seneca, Wyandotte, Ojibwe and many other Indigenous peoples. Specifically, the university resides on land ceded in the 1795 Treaty of Greeneville and the forced removal of tribes through the Indian Removal Act of 1830. As a land grant institution, we want to honor the resiliency of these tribal nations and recognize the historical contexts that has and continues to affect the Indigenous peoples of this land.

\software{\vice \citep{johnson_impact_2020}, Astropy \citep{astropy_collaboration_astropy_2013,astropy_collaboration_astropy_2018,astropy_collaboration_astropy_2022}, scikit-learn \citep{pedregosa_scikit-learn_2011}, SciPy \citep{virtanen_scipy_2020}, Matplotlib \citep{hunter_matplotlib_2007}, NumPy \citep{harris_array_2020}, pandas \citep{reback_pandas-devpandas_2021,the_pandas_development_team_pandas-devpandas_2025}, ChEAP \citep{palicio_analytic_2023}.}

\appendix

\begin{figure*}
    \centering
    \includegraphics[width=\textwidth]{\figpath{infall_parameters.pdf}}
    \caption{Gas abundance tracks in the [O/Fe]--[Fe/H] plane for one-zone chemical evolution models with different values for the infall parameters (see Section \ref{sec:infall-parameters}). In each panel, one parameter is varied according to the legend while the other two are held fixed. The open symbols along each curve mark points in time, as denoted in panels (b) and (e). The marginal panels show the corresponding stellar abundance distributions, which are convolved with a Gaussian kernel with a width of 0.02 dex for visual clarity. All models use the same fiducial parameters as the $R_{\rm gal}=8\kpc$ ring in the multi-zone models. The black lines represent the same set of parameters in each panel. The underlying 2-D histogram plots the APOGEE stellar abundance distribution in the Solar annulus ($7\leq R_{\rm gal}<9\kpc$, $|z|<2\kpc$), and the gray curves in the marginal panels plot the APOGEE 1-D abundance distributions. {\it Top row:} models adopting the \yZ{1} yields and $\eta=0.2$. {\it Bottom row:} models adopting the \yZ{2} yields and $\eta=1.4$.}
    \label{fig:infall-parameters}
\end{figure*}

\section{Reproducibility}
\label{app:reproducibility}

The figures, tables, and models in this paper are reproducible. The git repository associated to this study is publicly available at \url{\GitHubURL}, and the release {\tt v1.1.0} allows anyone to re-build the entire manuscript. The multi-zone model outputs used to produce the figures in this work are stored at \url{https://doi.org/10.5281/zenodo.16649938}.

\section{Infall Parameters}
\label{app:infall-parameters}

Figure \ref{fig:infall-parameters} illustrates the effect of the infall parameters $\tau_1$, $\tau_2$, and $t_{\rm max}$ (see Equation \ref{eq:twoinfall-ifr}) on gas abundance tracks in the [O/Fe]--[Fe/H] plane. Our fiducial parameters of $\tau_1=0.3\Gyr$, $\tau_2=15\Gyr$, and $t_{\rm max}=3.2\Gyr$ (for \yZ{1}) or $t_{\rm max}=2.2\Gyr$ (for \yZ{2}) are represented by the black curves. We detail the effects of these parameters and the reasons for our fiducial choices below.

Panels (a) and (d) illustrate the effect of varying the first infall timescale $\tau_1$. Shortening the timescale leads to more rapid enrichment along the high-$\alpha$ sequence, moving the knee to slightly higher metallicity. The gas abundance also reaches lower [O/Fe] and higher [Fe/H] at $t_{\rm max}$, although due to the low star formation rate few stars are formed in this regime. The effect of $\tau_1$ on the marginal [O/Fe] and [Fe/H] distributions is minimal. We adopt $\tau_1=0.3\Gyr$ for consistency with the literature \citep[e.g.,][]{nissen_high-precision_2020,palicio_analytic_2023,hegedus_reconstructing_2025}.

Panels (b) and (e) illustrate the effect of varying the second infall timescale $\tau_2$. The effect on the gas abundance tracks is relatively small compared to the other parameters, but the stellar abundance distributions are more strongly affected. Shortening the timescale means the star formation rate is higher immediately following $t_{\rm max}$, which enlarges the low-$\alpha$ loop. As a result, lower values of $\tau_2$ produce an [O/Fe] distribution that is more strongly tri-modal. This parameter varies with radius in the multi-zone models, and we adopt $\tau_2(R_\odot)=15\Gyr$ to minimize the width of the low-$\alpha$ peak in the Solar neighborhood.

Panels (c) and (f) illustrate the effect of varying the second infall onset time $t_{\rm max}$. A longer delay means the evolution proceeds farther down the high-$\alpha$ track before dilution, but conversely the low-$\alpha$ track does not extend as far. By affecting the end-point, the choice of $t_{\rm max}$ also shifts the MDF and [O/Fe] distributions slightly. For very short delays of $t_{\rm max}\sim1\Gyr$, the characteristic loop disappears entirely as the gas is still in the high-$\alpha$ regime when the second infall occurs. We adopt $t_{\rm max}=3.2\Gyr$ for most models, with $t_{\rm max}=2.2\Gyr$ for some \yZ{2} models to more closely match the analytic models of \citet{palicio_analytic_2023}.

Contrasting panels in the first and second row reveals the effect of changing the yield scale (see Table \ref{tab:yields}). With higher yields, enrichment proceeds more rapidly before the onset of SNe Ia, which moves the high-$\alpha$ knee to higher metallicity. After $t_{\rm max}$, the higher yield scale enhances O enrichment from CCSNe, which enlarges the low-$\alpha$ loop (see further discussion in Section \ref{sec:abundance-distributions}). We maintain a consistent end-point by scaling the outflow mass-loading factor $\eta_\odot$ with the yields.

\bibliography{references}
\bibliographystyle{aasjournalv7}

\end{document}